\documentclass[prb,twocolumn,showpacs]{revtex4}
\usepackage{amsmath}
\usepackage{graphicx}

\newcommand{\be}{\begin{equation}}
\newcommand{\ee}{\end{equation}}
\newcommand{\eps}{\varepsilon}
\newcommand{\mb}[1]{\mathbf{#1}}
\newcommand{\dl}{\widehat\Delta(\mb k)}
\newcommand{\nn}{\nonumber}

\newcommand{\Journal}[4]{#1 \textbf{#2}, #3 (#4)}

\begin{document}

\title{Mixed-parity superconductivity in centrosymmetric crystals}

\author{I. A. Sergienko}
\affiliation{Department of Physics and Physical Oceanography,
Memorial University of Newfoundland, St.\ John's, NL, A1B 3X7, Canada}

\begin{abstract}
A weak-coupling formalism for superconducting states possessing both singlet 
(even parity) and triplet (odd parity) components of the order parameter 
in centrosymmetric crystals is developed. It is shown that the quasiparticle 
energy spectrum may be non-degenerate even if the triplet component is unitary.
The superconducting gap of a mixed-parity state may have line nodes in the 
strong spin-orbit coupling limit. The pseudospin carried
by the superconducting electrons is calculated, from which follows a prediction 
of a kink anomaly in the temperature dependence of muon spin relaxation rate. 
The anomaly occurs at the phase boundary between the bare triplet
and mixed-parity states. The stability of mixed-parity states
is discussed within Ginzburg-Landau theory. The results may have immediate
application to the superconducting series Pr(Os$_{1-x}$Ru$_x$)$_4$Sb$_{12}$.
\end{abstract}


\maketitle

\section{Introduction}
Very recently, magnetic susceptibility and electrical resistivity measurements on 
the filled skutterudite series Pr(Os$_{1-x}$Ru$_x$)$_4$Sb$_{12}$ 
found that superconductivity exists in the whole concentration range
$0\leq x \leq 1$.~\cite{Frederick03b} While PrRu$_4$Sb$_{12}$ is a conventional
$s$-wave superconductor~\cite{Takeda00} with $T_c$=1 K, PrOs$_4$Sb$_{12}$ is a heavy
fermion material~\cite{Bauer02} with $T_c$=1.85 K. There is experimental evidence 
that PrOs$_4$Sb$_{12}$ has two superconducting (SC) phases~\cite{Bauer02,Vollmer03,
Izawa03} and that the quasiparticle energy spectrum has point 
nodes.\cite{Izawa03,Chia03} These facts indicate that the SC order 
parameter is definitely unconventional. Zero-field muon-spin relaxation ($\mu$SR) 
measurements suggest that the pairing in 
PrOs$_4$Sb$_{12}$ can be triplet (odd parity).\cite{Aoki03b}
The competition between the conventional $s$-wave and triplet
order parameters may result in the appearance of a new SC 
state in Pr(Os$_{1-x}$Ru$_x$)$_4$Sb$_{12}$ which has both singlet and triplet 
components and therefore is a mixed-parity (MP) state. 

So far, several systems in which MP SC states 
may occur have been studied theoretically. Mineev and Samokhin\cite{Mineev94} 
showed that the normal state may be unstable with respect to a helical MP structure
if the product of the two different parity representations of the symmetry group
contains a vector 
representation. The possible 
occurrence of MP states due to inhomogeneity of the order parameter in the 
Larkin-Ovchinnikov-Fulde-Ferrel state was addressed in 
Refs.~\onlinecite{Matsuo94, Shimahara00}. MP states in two-dimensional 
superconductors, in which inversion symmetry is broken due to low dimensionality, 
were studied by Gor'kov and Rashba.\cite{Gorkov01} Finally, an intriguing 
prediction was made by Volkov \emph{et al}.\cite{Volkov03,Bergeret03} about the 
possibility of generating a triplet condensate in mesoscopic ferromagnet/singlet 
superconductor multilayers. 

In this paper, we analyse MP SC states in \emph{bulk crystals with 
inversion symmetry} in zero magnetic field and describe the possible experimental 
manifestation of the MP state in Pr(Os$_{1-x}$Ru$_x$)$_4$Sb$_{12}$. 
We discuss the possibility of the realization of a MP state in terms 
of Ginzburg-Landau (GL) theory. We develop a weak-coupling
formlaism for MP states based on the generalized BCS approach.\cite{Sigrist91}
We obtain the quasiparticle energy spectrum and show that it may be non-degenerate 
even if the triplet component is \emph{unitary}. It is shown that the gap function
of the MP state in the strong spin-orbit limit may have \emph{line nodes}, in 
contrast to bare triplet states. The Gor'kov formalism is used
to obtain the self-consistent equation of the gap function and to calculate the 
pseudospin carried by the SC electrons. A kink anomaly in the temperature dependence 
of the muon spin relaxation rate is expected on the boundary between the bare triplet 
and MP states.

\section{\label{SecGL}Ginsburg-Landau model}
The explicit 
form of the GL functional can be established only if the transformation properties of 
the multidimensional order parameters with respect to the spatial symmetry 
transformations are known. Thus far, the symmetry of the order parameter in
PrOs$_4$Sb$_{12}$ has not been determined unambiguously.\cite{Sergienko03} 
Also, the effects of multi-dimensionality of the order parameter are beyond the scope 
of this paper. Thus, we adopt an effective one-component model which only takes into 
account gauge, inversion and time reversal but not the crystallographic 
symmetry. We denote by $\eta=|\eta|e^{i\phi_1}$ the non-vanishing component of the 
singlet order parameter and by $\xi=|\xi|e^{i\phi_2}$ that of the triplet order 
parameter. 
A homogeneous situation in zero magnetic field is considered in the following, so that
the gradient terms of $\eta$ and $\xi$ are neglected. The GL potential is
\begin{eqnarray}\label{pot}
F&=&a_1[T-T_{c1}(x)]\, |\eta|^2 + b_1[T-T_{c2}(x)]\, |\xi|^2\\
&& + a_2 |\eta|^4+ b_2 |\xi|^4 + c_1 |\eta|^2 |\xi|^2 + 
c_2 (\eta^2 \xi^{*2}+ \eta^{*2} \xi^2),\nn
\end{eqnarray}
where $T$ is temperature and $T_{c1,2}(x)$ are concentration dependent transition
temperatures for the singlet and triplet order parameters, respectively. The latter
can be roughly fitted to the experimental data\cite{Frederick03b} as linear functions 
$T_{c1}(x)=T_s [1-\alpha(1-x)]$, $T_{c2}(x)=T_t (1-\beta x)$ with $T_s=1.2$ K, 
$T_t=1.8$ K, $\alpha=1.07$, and $\beta=1.03$. The rest of the parameters 
in~(\ref{pot}) are assumed to be undetermined constants.\cite{ParityNote}

\begin{figure}[t]
\caption{\label{PD}Sketch of the phase diagram of Pr(Os$_{1-x}$Ru$_x$)$_4$Sb$_{12}$.
Solid lines represent a fit to the experimental data.\cite{Frederick03b}
The other lines are calculated from the model~(\ref{pot}).
(a) Broken lines are the boundaries of the MP state for $4a_2b_2>(c_1-2|c_2|)^2$
[See Eqs.~(\ref{lineSingl}) and~(\ref{lineTripl})]. 
(b) Dash-dot line represents a first-order phase boundary for 
$4a_2b_2<(c_1-2|c_2|)^2$ [Eq.~(\ref{lineFirst})].}
\includegraphics[width=8.5cm]{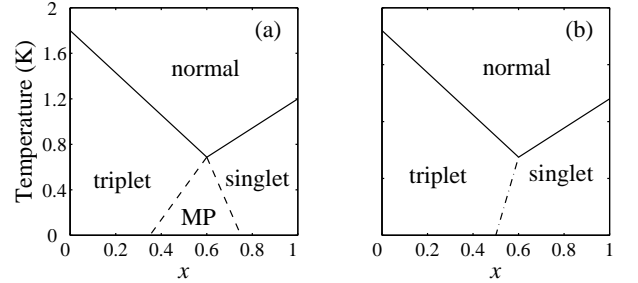}
\end{figure}

$F$ can be immediately minimized with respect to the relative phase 
$\Delta\phi=\phi_1-\phi_2$. Depending on the sign of $c_2$, the following MP states 
are energetically favoured
\begin{subequations}
\label{states}
\begin{eqnarray}
\Delta\phi=\pi/2, 3\pi/2, & \eta\xi^*=\pm i|\eta||\xi| & \text{for } c_2 > 0; 
\label{stateA}\\
\Delta\phi=0, \pi,\qquad & \eta\xi^*=\pm|\eta||\xi|& \text{for } c_2 < 0.
\label{stateB}
\end{eqnarray}
\end{subequations}
The phase diagram of~(\ref{pot}) adapted for Pr(Os$_{1-x}$Ru$_x$)$_4$Sb$_{12}$ 
is shown in Fig.~\ref{PD}. If $4a_2b_2>(c_1-2|c_2|)^2$, one of the MP 
states~(\ref{states}) exists in the sector bounded by two second-order transition 
lines 
\begin{equation}\label{lineSingl}
T(x)=\frac{c\alpha_1 T_{c1}(x)-2a_2\beta_1T_{c2}(x)}{c\alpha_1-2a_2\beta_1}
\end{equation}
on the singlet side and 
\begin{equation}\label{lineTripl}
T(x)=\frac{2b_2\alpha_1 T_{c1}(x)-c\beta_1T_{c2}(x)}{2b_2\alpha_1-c\beta_1}
\end{equation}
on the triplet side. If $4a_2b_2<(c_1-2|c_2|)^2$, the MP states~(\ref{states}) 
do not minimize $F$ and a first-order phase transition between the bare 
singlet and triplet states occurs at the line 
\begin{equation}\label{lineFirst}
T(x)=\frac{\alpha_1\sqrt{b_2} T_{c1}(x)-\beta_1\sqrt{a_2}T_{c2}(x)}{
\alpha_1\sqrt{b_2}-\beta_1\sqrt{a_2}}.
\end{equation}

Phenomenologically, therefore, both types of the phase diagram shown in 
Fig.~\ref{PD} may occur with the same probability in the sense that they are described
by the same fourth-order model~(\ref{pot}). Another doped system 
U$_{1-x}$Th$_x$Be$_{13}$ is an example in which the co-existence of two SC order 
parameters is preferred to the first-order phase transition scenario.
\cite{Sigrist89,UBeNote}

\section{\label{SecBCS}Generalized BCS approach}

A realistic description of superconductivity in Pr(Os$_{1-x}$Ru$_x$)$_4$Sb$_{12}$ 
should be based on a strong-coupling approach which would incorporate the 
superconductivity mechanism and coupling of SC electrons to the Os(Ru) ions.
However, at present even the superconductivity mechanism of pure PrOs$_4$Sb$_{12}$ 
remains vague. In this paper our goal is to elucidate the qualitative features of MP 
states rather than to make quantitative predictions. Hence, we use the weak-coupling 
approach, implicitly assuming that the main ingredients 
of the theory, the band energy relative to the chemical potential $\eps(\mb k)$ and 
the pairing interaction matrix element $V_{s_1 s_2 s_3 s_4}(\mb k, \mb k')$, depend 
on the Ru concentration $x$. 

The generalized mean-field BCS Hamiltonian is~\cite{Sigrist91}
\be
\begin{array}{l}\label{ham}
\displaystyle
H=\sum_{\mb k, s} \eps(\mb k) a^\dagger_{\mb k s}a_{\mb k s}
+\frac1 2 \sum_{\mb k, s_1, s_2} [
\Delta_{s_1 s_2}(\mb k)a^\dagger_{\mb k s_1}a^\dagger_{-\mb k s_2}\\
\qquad\qquad\qquad\qquad\qquad\qquad\quad
-\Delta^*_{s_1 s_2}(-\mb k)a_{-\mb k s_1}a_{\mb k s_2}],
\end{array}
\ee
where $a^\dagger_{\mb k s}$ ($a_{\mb k s}$) is the creation (annihilation) operator of
an electron with wave vector $\mb k$ and pseudospin\cite{Ueda85} $s$ and the gap 
function matrix is
\be\label{delDef}
\Delta_{s_1 s_2}(\mb k)=-\sum_{\mb k', s_3, s_4} V_{s_2 s_1 s_3 s_4}
(\mb k, \mb k') \langle a_{\mb k' s_3}a_{-\mb k' s_4} \rangle.
\ee
The gap function is conveniently parametrized by an even scalar 
function $\psi(\mb k)=\sum_i \eta_i\psi_i(\mb k)$ for the singlet component and an odd 
vectorial function $\mb d(\mb k)=\sum_i \xi_i\mb d_i(\mb k)$ for the triplet 
component. 
Here $\psi_i(\mb k)$ and $\mb d_i(\mb k)$ are basis functions of the irreducible
representations of the point group and $\eta_i$ and $\xi_i$ are the corresponding order
parameters.\cite{Sigrist91}  In the MP state the gap function takes the form
\begin{equation}
\dl=[\psi(\mb k)\widehat\sigma_0 + \mb d(\mb k) \widehat{\boldsymbol{\sigma}}]
i \widehat\sigma_y,
\end{equation}
where $\widehat\sigma_0$ is the $2\times2$ unit matrix and $\widehat{\boldsymbol
{\sigma}}=(\widehat\sigma_x, \widehat\sigma_y, \widehat\sigma_z)$ are the Pauli 
matrices.

The Bogoliubov transformation for the diagonalization 
of~(\ref{ham}) yields the quasiparticle energy spectrum
\be\label{spectr}
 E_{\mb k \pm}=\sqrt{\eps(\mb k)^2 + \Delta_\pm(\mb k)^2},
\ee
with two gaps $\Delta_\pm(\mb k)$ defined by
\be\label{gap2}
\Delta_\pm(\mb k)^2 = |\psi(\mb k)|^2 + |\mb d(\mb k)|^2 \pm |\mb p(\mb k) + 
\mb q(\mb k)|,
\ee
where two real vectors $\mb p(\mb k) = \psi (\mb k)\mb d^*(\mb k) + \psi^*(\mb k)
 \mb d(\mb k)$ and $\mb q(\mb k)=i[\mb d(\mb k) \times \mb d^*(\mb k)]$ are introduced.
They are orthogonal to each other, 
therefore 
\begin{equation}
|\mb p(\mb k) + \mb q(\mb k)|=\sqrt{\mb p(\mb k)^2+ \mb q(\mb k)^2}.
\end{equation} 
Note that if $\mb q(\mb k)\ne 0$, \emph{i. e.} $\mb d(\mb k)$ and $\mb d^*(\mb k)$ are 
not collinear, 
then $\mb p(\mb k)$ is finite in the MP state. 

It follows that the quasiparticle spectrum~(\ref{spectr}) 
is degenerate only if $\mb p (\mb k)= \mb q(\mb k)=0$. Such a state corresponds 
to~(\ref{stateA}).
It is not invariant with respect to time reversal $\cal K$, but it possesses a 
combined symmetry element $U_1(\pi)I \cal K$, where $U_1$ is the gauge transformation 
and $I$ is inversion. 
In a time reversal invariant MP state, $\mb q(\mb k)=0$ and 
$p(\mb k)=2|\psi(\mb k)||\mb d(\mb k)|$ [See~(\ref{stateB})]. Then, the gap acquires 
the form 
\be\label{eq8}
\Delta_\pm(\mb k)^2=(|\psi(\mb k)| \pm |\mb d(\mb k)|)^2. 
\ee

\begin{figure}[t]
\caption{\label{nodes}(Colour online) Line nodes in the SC gap of the 
time-reversal invariant admixture of $s$- and $p$-wave states~(\ref{components}) for 
$|\xi_1|>|\xi_2|$. (a) $|\eta| < |\xi_2|$; (b) $|\eta| = |\xi_2|$; (c) 
$|\xi_2| < |\eta| < |\xi_1|$.}
\includegraphics[width=8.5cm]{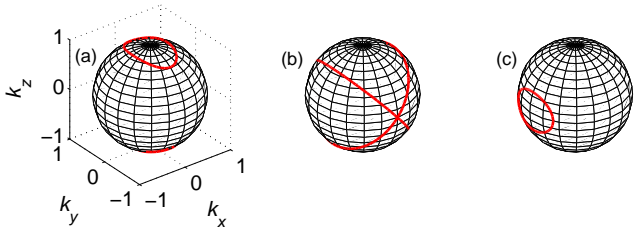}
\end{figure}

As follows from~(\ref{gap2}), the gap of a MP state vanishes if both singlet and 
triplet components have nodes, \emph{i.~e.}  $\psi(\mb k)=0$ and $|\mb d(\mb k)|^2
=|\mb d(\mb k) \times \mb d^*(\mb k)|$. 
In the strong spin-orbit coupling limit,\cite{HarimaNote} these 
nodes can be found only at isolated points on the Fermi 
surface~\cite{Volovik85,Blount85} unless 
the Fermi surface crosses the boundary of the Brillouin zone.~\cite{Norman95}
However, $\Delta_-$ may have additional \emph{line nodes} 
if $\mb p \ne 0$, as illustrated by the following example. 

Keeping in mind the possible application to Pr(Os$_{1-x}$Ru$_x$)$_4$Sb$_{12}$, 
we consider a $p$-wave state which may arise under $T_h$ crystallographic 
symmetry. Time reversal symmetry is broken in the SC state of 
PrOs$_4$Sb$_{12}$.\cite{Aoki03b} Here, for the sake of simplicity, a time-reversal 
invariant admixture of $s$- and $p$-wave states is considered, in which the gap 
takes the simple form~(\ref{eq8}). We can show that this simplification does not affect
the main result, line nodes may exist for finite $\mb q$ as well.

Let us take a $p$-wave state belonging to the irreducible representation $T_u$ of
$T_h$. This representation has two linearly independent sets of basis 
functions.\cite{Sergienko03} Therefore, in general, one should take into account two 
order parameters of the same symmetry. We consider the state denoted $(0, 0, 1)$ 
in the order parameter space.\cite{Sergienko03} The MP gap function
is given by 
\be\label{components}
\psi (\mb k)=\eta, \qquad 
\mb d(\mb k)=\xi_1 k_x \widehat \mb y + \xi_2 k_y \widehat \mb x, 
\ee
where $\eta$, $\xi_1$ and $\xi_2$ are the $T$ and $x$ dependent order parameters. 
$\Delta_-(\mb k)$ vanishes at two lines defined by the equation 
\be
|\xi_1|^2 k_x^2 + |\xi_2|^2 k_y^2 = |\eta|^2.
\ee
The line nodes exist 
while $|\eta|\le \max(|\xi_1|, |\xi_2|)$. This condition is  
obeyed close to the boundary with the bare triplet state. Fig.~\ref{nodes} shows the
line nodes on a unit spherical Fermi surface for $|\xi_1|>|\xi_2|$. Note that in 
this example the line nodes are \emph{not} remnants of line nodes of the singlet 
state, which is fully gapped, but rather they appear due to a distortion of point 
nodes of the triplet component.

\section{\label{SecGreen}Green's functions and gap equation}

Beginning with the equations of motion for Heisenberg operators, 
the Gor'kov equations\cite{Gorkov58} for the normal $G_{s s'}(\mb k, \tau)$ and 
anomalous 
$F_{s s'}(\mb k, \tau)$, $F^\dagger_{s s'}(\mb k, \tau)$ temperature Green's 
functions in the MP state are obtained in the usual way.\cite{Abrikosov63,Mahan90} 
The Green's functions are defined as
\begin{eqnarray}\label{defGreen}
G_{s s'}(\mb k, \tau)&=&-\langle T_\tau\{a_{\mb k s}(\tau)a^\dagger_{\mb k s'}
(0)\} \rangle \nn\\ 
F_{s s'}(\mb k, \tau)&=&\langle T_\tau\{a_{\mb k s}(\tau)a_{-\mb k s'}(0)\}
\rangle \nn\\
F^\dagger_{s s'}(\mb k, \tau)&=&\langle T_\tau\{a^\dagger_{-\mb k s}(\tau)
a^\dagger_{\mb k s'}(0)\}\rangle,
\end{eqnarray}
where $T_\tau$ is the imaginary time ordering operator. The Fourier transform is
$ \widehat A(\mb k, \tau) = T \sum_n \widehat A(\mb k, \omega_n) e^{-i\omega_n \tau}$,
where $\widehat A$ stands for $\widehat G$, $\widehat F$, or $\widehat F^\dagger$;
$T$ is temperature and $\omega_n=\pi T (2n+1)$ is the Matsubara frequency for
fermions. 

The resulting Gor'kov equations for the Fourier transforms are
\begin{eqnarray}
[i\omega_n-\eps(\mb k)]\widehat G(\mb k, \omega_n)+  
\dl \widehat F^\dagger(\mb k, \omega_n) &=&\widehat \sigma_0,\nn\\ 
\left[i\omega_n+\eps(\mb k)\right]
\widehat F^\dagger(\mb k, \omega_n)+
\widehat\Delta^\dagger(\mb k) \widehat G(\mb k, \omega_n) &=& 0.
\end{eqnarray}
They are solved by 
\begin{widetext}
\begin{eqnarray}
\widehat G(\mb k, \omega_n)&=&\frac{
-[\omega_n^2+\eps(\mb k)^2+|\psi(\mb k)|^2+|\mb d(\mb k)|^2]\widehat\sigma_0
+[\mb p(\mb k) + \mb q(\mb k)]\widehat{\boldsymbol\sigma}}
{(\omega_n^2+E_{\mb k +}^2)(\omega_n^2+E_{\mb k -}^2)}[i\omega_n+\eps(\mb k)]
,\nn\\
\widehat F(\mb k, \omega_n)&=&\frac{[\omega_n^2+\eps(\mb k)^2+|\psi(\mb k)|^2
+|\mb d(\mb k)|^2] \dl - |\mb p(\mb k)+\mb q(\mb k)|\widehat\Omega(\mb k)}
{(\omega_n^2+E_{\mb k +}^2)
(\omega_n^2+E_{\mb k -}^2)}, \label{solGreen}
\end{eqnarray}
\end{widetext}
where 
\be
\widehat \Omega (\mb k)
= \frac{\mb d(\mb k) \mb p(\mb k) \widehat\sigma_0 + [\psi(\mb k)\mb p(\mb k)
+i \mb q(\mb k)\times \mb d(\mb k)]
\widehat{\boldsymbol \sigma}}{|\mb p(\mb k)+\mb q(\mb k)|}(i\widehat\sigma_y).
\ee 
The denominator
introduced in the definition of $\widehat \Omega(\mb k)$
allows one to put the self-consistent gap equation in a concise form. 
Using~(\ref{delDef}), (\ref{defGreen}), 
and (\ref{solGreen}) one 
obtains
\be
\Delta_{s_1 s_2}(\mb k)=- \sum_{\mb k', s_3, s_4} 
V_{s_2 s_1 s_3 s_4}(\mb k, \mb k') {\cal F}_{s_3 s_4}(\mb k', T),
\ee
where
\begin{eqnarray}
\widehat{\cal F}(\mb k, T)
&=& \frac{\dl+\widehat\Omega(\mb k)}{4 E_{\mb k +}} 
\tanh \left(\frac{E_{\mb k +}}
{2T}\right)\nn\\
&& + \frac{\dl-\widehat \Omega(\mb k)}{4 E_{\mb k -}} 
\tanh \left(\frac{E_{\mb k -}}
{2T}\right).
\end{eqnarray}

\section{\label{SecMSR}Muon spin rotation in Mixed-Parity states}
$\mu$SR measurements are widely used to reveal the nature of a SC 
state as they provide invaluable information about the distribution of the local 
magnetic field $\mb H$. There are several possible sources of internal magnetic 
fields in superconductors. They include magnetic moments of localized states, 
two-dimensional imperfections such as domain walls and the 
surface,\cite{Volovik85,Sigrist91} and magnetic moment of 
Cooper pairs. Sigrist and Rice~\cite{Sigrist89} showed that the latter 
contribution always vanishes in singlet pairing states, and it is non-vanishing in 
bare triplet states only if $\mb q(\mb k)\ne 0$. 
The zero-field $\mu$SR spectra are usually fitted using of the Kubo-Toyabe 
function~\cite{Hayano79}
\be
g(t)=\frac 1 3+\frac 23(1-\delta^2 t^2)\exp(-\frac 1 2 \delta^2 t^2),
\ee 
where $t$ is time and the relaxation rate $\delta$ is 
proportional to $\sqrt{\langle \mb H^2 \rangle}$.
The temperature dependence of $\delta$ 
in the MP states can be obtained using the Green's functions~(\ref{solGreen}).

A calculation of the actual magnetic moment with strong spin-orbit coupling requires 
detailed knowledge of the electronic band structure. At every point $\mb k$ in
 momentum space, the operators $a_{\mb k s}^\dagger$ and $a_{\mb k s}$ can
be expressed as linear combinations of Bloch sums consisting of orbital 
and spin parts. In the calculation of the magnetic moment, a quadratic form of the 
coefficients of this transformation will be an additional factor in the summand of 
the resulting 
expression~(\ref{spin}) (see below). However, the qualitative features of 
$\delta(T)$ may be discerned by calculating the average pseudospin of the 
Cooper pairs.\cite{Ueda85} The pseudospin operator is defined as
\begin{eqnarray}
S_x &=& \dfrac 12 \sum_{\mb k}(a_{\mb k \Uparrow}^\dagger a_{\mb k \Downarrow}+
a_{\mb k \Downarrow}^\dagger a_{\mb k \Uparrow}),\nn\\
S_y &=& \dfrac i2 \sum_{\mb k}(-a_{\mb k \Uparrow}^\dagger a_{\mb k \Downarrow}+
a_{\mb k \Downarrow}^\dagger a_{\mb k \Uparrow}),\nn\\
S_z &=& \dfrac 12 \sum_{\mb k}(a_{\mb k \Uparrow}^\dagger a_{\mb k \Uparrow}-
a_{\mb k \Downarrow}^\dagger a_{\mb k \Downarrow}),\label{defspin}
\end{eqnarray}
where the indices $\Uparrow$ and $\Downarrow$ denote the pseudospin-up and 
down states, respectively. Using Eqs.~(\ref{defGreen}), 
(\ref{solGreen}), (\ref{defspin}), and the oddness of $\mb p(\mb k)$,
one obtains the average pseudospin of the Cooper pairs,
\be
\begin{array}{rl}\label{spin}
\displaystyle \mb S (T) = \sum_{\mb k} 
\dfrac{\eps(\mb k)\,\mb q(\mb k)}{|\mb p(\mb k) + \mb q(\mb k)|} & \left[
\dfrac{\tanh(E_{\mb k-}/2T)}{4E_{\mb k-}}\right.\\
& \quad \left. - \dfrac{\tanh(E_{\mb k+}/2T)}{4E_{\mb k+}}\right].
\end{array}
\ee
Hence, as in the case of a bare triplet state, $\mb S(T)$ vanishes if 
$\mb q(\mb k) = 0$.
If the bare triplet state is nonunitary [$\mb q(\mb k)\ne 0$] then, as follows 
from~(\ref{spin}), the temperature dependence of $\delta$ has a kink at 
the point of the bare triplet-to-MP phase transition with a discontinuity in 
$d\delta/d T$ proportional to $d \mb p(\mb k)^2/d T$.

\section{\label{SecSum}Summary} Using a GL-type model, we showed that a MP SC state
can be stable in a finite region of the $T$-$x$ phase diagram of 
Pr(Os$_{1-x}$Ru$_x$)$_4$Sb$_{12}$. 
The phase transitions from the bare singlet
and triplet states to the MP state are second-order.
The general formalism describing the MP states within BCS 
approach is developed. The quasiparticle energy spectrum is degenerate
only if $\mb p(\mb k)=\mb q(\mb k) = 0$. The gap in the MP state can have line nodes in 
strong spin-orbit coupling limit if $\mb p(\mb k) \ne 0$. Normal and anomalous Green's 
functions and the self-consistent gap equation are obtained. 
The MP state may be experimentally indicated by line nodes 
in the gap of the quasiparticle spectrum and a kink in $\mu$SR relaxation rate
$\delta(T)$.

\begin{acknowledgments}
I thank S. H. Curnoe, Y. Aoki, and H. Harima for valuable discussions and 
K. V. Samokhin for bringing my attention to Ref.~\onlinecite{Mineev94}.
This work was supported by NSERC Canada. 
\end{acknowledgments}


\begin{thebibliography}{00}
\bibitem{UBeNote} The two order parameters in U$_{1-x}$Th$_x$Be$_{13}$ have the same 
parity.
\bibitem{HarimaNote} Although $4f$ electrons of Pr in PrOs$_4$Sb$_{12}$ are 
localized,~\cite{Sugawara02} spin-orbit coupling strongly affects the Fermi surface
due to $5d$ electrons of Os and $5p$ electrons of Sb (H. Harima, private 
communication).
\end{thebibliography}
\end{document}